\documentclass[9pt,twocolumn,twoside]{osajnl}

\journal{ol} 

\setboolean{shortarticle}{true}

\title{Ultra-efficient and highly-tunable second-harmonic generation in Z-cut periodically poled lithium niobate nanowaveguides}
\author[1,2]{Jia-Yang Chen}
\author[1,2]{Chao Tang}
\author[1,2]{Zhao-Hui Ma}
\author[1,2]{Zhan Li}
\author[1,2]{Yong Meng Sua}
\author[1,2,*]{Yu-Ping Huang}

\affil[1]{Department of Physics, Stevens Institute of Technology, 1 Castle Point Terrace, Hoboken, NJ 07030, USA}
\affil[2]{Center for Quantum Science and Engineering, Stevens Institute of Technology, 1 Castle Point Terrace, Hoboken, NJ 07030, USA}

\affil[*]{Corresponding author: yuping.huang@stevens.edu}

\begin{abstract}
Thin-film lithium niobate on insulator (LNOI) has emerged as a superior integrated-photonics platform for linear, nonlinear, and electro-optics. Here we combine quasi-phase-matching, dispersion engineering, and tight mode confinement to realize nonlinear parametric processes with both high efficiency and wide wavelength tunability. On a millimeter-long, Z-cut LNOI waveguide, we demonstrate ultra-efficient ($1900\pm500 \% $W$^{-1}$cm$^{-2}$) and highly tunable (-1.71 nm/K) second-harmonic generation from 1530 to 1583 nm by type-0 quasi-phase-matching. Our technique is applicable to optical harmonic generation, quantum light sources, frequency conversion, and many other photonic information processing across visible to mid-IR spectral bands.
\end{abstract}

\setboolean{displaycopyright}{true}

\begin{document}

\maketitle

Efficient second-order nonlinear ($\chi^{(2)}$) processes are desirable for a plethora of applications such as nonclassical light sources \cite{guo2017parametric, luo2017chip, Chen:19,Elkus:19}, optical frequency conversion \cite{luo2018highly, Wang:18PPLN, Rao:19, niu2020optimizing, Stanton:20}, low-threshold optical parametric oscillators \cite{Bruch:19}, and supercontinuum generation \cite{Lu:s, Jankowski:20}. In particular, the frequency conversion can flexibly create coherent light at needed wavelengths not easily accessible by laser gain media. It enables numerous optical studies and applications in spectroscopy, frequency metrology, remote sensing, etc., where coherent light at certain specific wavelengths is essential. 
Among others, second-harmonic generation (SHG) has long been deployed in a multitude of laser-based applications, which would benefit significantly from high efficiency and tunability.

To this end, thin-film lithium niobate on insulator (LNOI) has been explored first to enhance the SHG efficiency \cite{Wang:18PPLN, Chen:19, Rao:19, niu2020optimizing}, and more recently to increase its wavelength tunability by thermo-optics \cite{luo2018highly}. To achieve high tunability, type-I phase matching scheme was adopted to take advantage of the largest thermo-optical birefringence coefficient in lithium niobate \cite{fejer1992quasi, luo2018highly}. However, such a scheme takes on a 30-fold reduction in the SHG efficiency, as it uses LNOI's $d_{31}$ nonlinear susceptibility whose typical value of 4.7 pm/V is much lower than that of $d_{33}$ at $\sim$27 pm/V. On the other hand, with periodic poling, a type-0 quasi-phase matching (QPM) scheme can utilize $d_{33}$ for high efficiency \cite{Sua:18}, but suffers a lower tunability because of a much small thermo-optic coefficient. This unfavorable trade-off between the efficiency and tunability poses a limitation in achieving higher performance of nonlinear optical devices made of lithium-niobate. 

Here, we demonstrate a new approach to overcome the trade off by careful group velocity engineering followed by quasi-phase matching. Specifically, we first minimize the group velocity mismatch (GVM) between the interacting wavelengths and then use appropriate periodic poling to compensate for their phase mismatching. This permits to sustain the high conversion efficiency using $d_{33}$ while enhancing the QPM tunability at the same time. With a dispersion engineered PPLN waveguide on a Z-cut LNOI chip,  we achieve ultra-efficient ($1900\pm500 \% $W$^{-1}$cm$^{-2}$) and highly tunable (-1.71 nm/K) SHG from 1530 to 1583 nm. Uniquely, the QPM wavelength exhibits blue shifts at a large tunability as the temperature is increased. Meanwhile, the phase matching profiles remain in a near-ideal $sinc^2$ shape and maintain high peak conversion efficiency over the entirely tuning range.   

The key to the current large tunability is the GVM reduction via the modal dispersion engineering, using optimized waveguide cross sections. Then, QPM is achieved by perodic poling to offset the phase mismatch caused by both the material and resulted modal dispersion. It is this two-step optimization that gives rise to the high efficiency and large tunability at the same time. Our technique is thus in contrast to the modal phase matching approach \cite{luo2018highly}, where the refractive indices of fundamental and second-harmonic modes need to be matched at the first place, leaving little room for the group velocity engineering. It is also not reproducible in bulky PPLN waveguides \cite{fejer1992quasi, Sua:18}, whose modal dispersion is dominated by the material dispersion due to a much larger cross section (e.g., $>$3$\times$3 $\mu m$), thus prohibiting the proposed GVM maneuvering.  

The paper is structured as follows. In section 1, we present the theory and simulation to study the thermal effects on GVM, followed by a brief description of the fabrication procedure. In section 2, we characterize the linear and nonlinear optical properties and thermo-optic tunablity of the LNOI waveguide. We finally conclude in Section 3.

\section{Design and Fabrication}
\subsection{Theory and simulation}
 Taking modal dispersion and thermo-optic effects into account, i.e. $n_1(\omega_1,T)$ and $n_2(\omega_2,T)$, the QPM condition of the PPLN waveguide reads \cite{Wang:18PPLN, Chen:19}
\begin{equation}
\Delta K= n_{2}(\omega_2,T) - n_{1}(\omega_1,T) -\frac{2\pi c}{\Lambda \omega_2} = 0,
    \label{eq2}
\end{equation}
where $\Lambda$ is the poling period, $n_1$ and $n_2$ are the effective refractive indices for fundamental ($\omega_{1}$) and SH ($\omega_{2}$) modes, respectively.
To derive its thermal dependency, we perform the Taylor's series expansion for each term in Eq.~(\ref{eq2}) around one of their perfect QPM points $\omega^o_1$, $\omega^o_2$, and $T^o$ \cite{luo2018highly}:
\begin{align}  \label{eq3}
 n_{2}(\omega_2,T)= n_{2}(\omega_2^o,T^o)+\frac{\partial n_2}{\partial \omega_2}\Delta \omega_2+\frac{\partial n_2}{\partial T}\Delta T, \nonumber\\
 n_{1}(\omega_1,T)= n_{1}(\omega_1^o,T^o)+\frac{\partial n_1}{\partial \omega_1}\Delta \omega_1+\frac{\partial n_1}{\partial T}\Delta T, \\
 \frac{2\pi c}{\Lambda \omega_2} = \frac{2\pi c}{\Lambda \omega_2^o}\left(1-\frac{\Delta \omega_2}{\omega_2^0}\right),\nonumber
 \end{align}
where $\omega_2 = 2\omega_1$, $\Delta \omega_2 = 2\Delta \omega_1$ and $n_{2}(\omega_2^o,T^o)-n_{1}(\omega_1^o,T^o)=\frac{2\pi c}{\Lambda \omega_2^o}$. After temperature is perturbed, the new system still need to fulfill QPM condition with the same poling period (thermal expansion is neglected here). By inserting the expanded terms and plugging individual items in Eq.(\ref{eq3}) into Eq.(\ref{eq2}) and simplifying with initial QPM condition of $\Delta K=0$, we arrive at :
\begin{eqnarray}
 \frac{\Delta \omega_1}{\Delta T} = \frac{\frac{\partial n_1}{\partial T}-\frac{\partial n_2}{\partial T}}{2\frac{\partial n_2}{\partial \omega_2}-\frac{\partial n_1}{\partial \omega_1}+\frac{4\pi c}{\Lambda \omega_2^{o2}}}.
    \label{eq4}
\end{eqnarray}

From the above equation, the temperature dependency of the phase matching comes from two folds. The first is exhibited in the numerator, given by the difference of thermo-optic coefficient between fundamental and SH modes. The second is in the denominator, governed by the group indices and an extra constant term. By increasing the numerator and/or reducing the denominator, we could increase the temperature tunability of the LNOI waveguide. For lithium niobate, $\frac{\partial n_e}{\partial T} \gg \frac{\partial n_o}{\partial T}$, \cite{TO2005}, so that one could use the fundamental-frequency mode along the ordinary axis (o-polarized) and the SH mode along the extraordinary axis (e-polarized), in order to maximize the thermal coefficient difference  \cite{luo2018highly}. However, it works at the expense of reducing the conversion efficiency over 30 times due to the much smaller nonlinear tensor $d_{31}$ it can support. This motivates us to explore another possibility, by minimizing the denominator. 

By recognizing $n_g = c/v_g=n+\omega \frac{\partial n}{\partial \omega}$, the group velocity mismatch GVM $= 1/v_{g1}-1/v_{g2}$, and $n_{2}^o-n_{1}^o=\frac{2\pi c}{\Lambda \omega_2^o}$, Eq.(\ref{eq4}) is reduced to:
\begin{equation}
 \frac{\Delta \omega_1}{\Delta T} = \frac{\frac{\partial n_2}{\partial T}-\frac{\partial n_1}{\partial T}}{GVM\frac{c}{\omega_1}}.
    \label{eq5}
\end{equation}
Hence, GVM can serve as an efficient knob to control the tunability, both for its magnitude and the direction of the QPM wavelength shifting. Meanwhile, the largest nonlinear tensor $d_{33}$ in lithium niobate can be accessed, thanks to the periodic poling. This method addresses the inefficiency drawback in Ref.~\cite{luo2018highly}.

\begin{figure}
\centering
\includegraphics[width=3in]{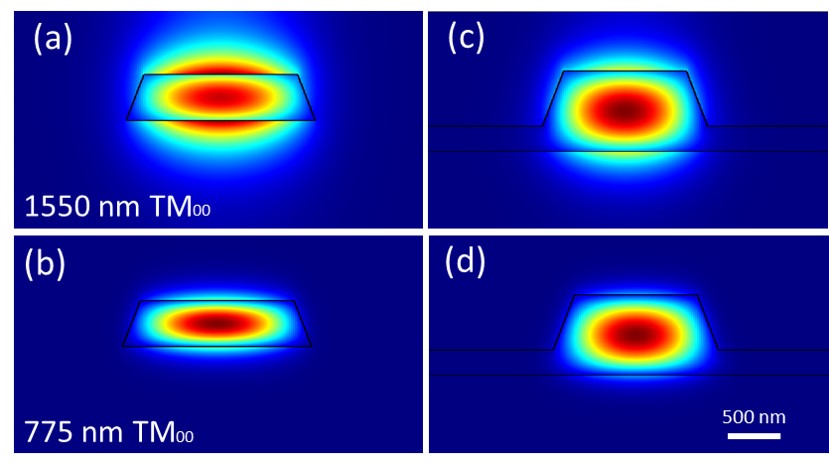}
\caption{Simulated profiles for the 1550-nm and 775-nm quasi-TM$_{00}$ modes. The waveguide cross section 400$\times$1850nm (height and top) for (a) and (b), and 700$\times$1500nm for (c) and (d). For latter one, the etched depth is 480 nm, so that a layser of 220 nm LN slab is remained. The simulated sidewall angle in is 62$^\circ$ for all figures.} 
\label{fig1}
\end{figure}

\begin{figure}
\centering
\includegraphics[width=3.4in]{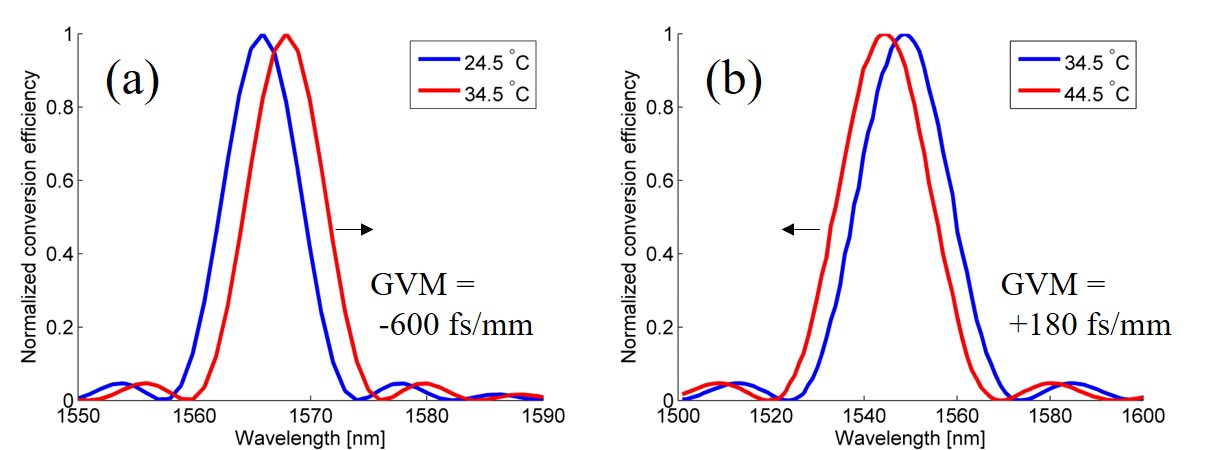}
\caption{Simulated phase matching curves of  Waveguide 1 (a) and Waveguide 2 (b) under different temperatures. Waveguide 1 has a negative GVM (-600 fs/mm) and red shifts at 0.2 nm/K. Waveguide 2 has a positive GVM (180 fs/mm) and blue shifts at 0.6 nm/K.} 
\label{fig2}
\end{figure}

We validate the theoretical prediction through performing MODE simulations of optical modes and their temperature dependency  (Lumerical, Inc). As a proof of principle, we compare the thermal effects on the GVM for two LNOI waveguide geometries (Waveguide 1: 400$\times$1850 nm and Waveguide 2: 700$\times$1500 nm). As shown in Fig.~\ref{fig1}, we select  fundamental quasi-transverse-magnetic (quasi-TM) modes for both 1550 nm and 775 nm in Z-cut LNOI waveguides, which allows to access their largest nonlinear tensor $d_{33}$ while maximizing the mode overlapping. With the chosen geometry, the poling period for type-0 phase matching is calculated to be 2.45 $\mu m$ and 3.8 $\mu m$, respectively. Later, by incorporating a temperature-dependent Sellmeier equation into the Lumerical simulator, we calculate their phase matching curves at various temperatures. As shown in Fig.~\ref{fig2}, Waveguide 1 exhibits red shift with a 0.2 nm/K slope. However, Waveguide 2 exhibits blue shifting with 0.6 nm/K slope, which shows opposite temperature dependency than in typical cases \cite{luo2018highly, Wang:18PPLN, Chen:19, fejer1992quasi}. This distinct behavior indicates that they have different signs in GVM. From these numbers we extract their GVM to be -600 fs/mm and 180 fs/mm, respectively. Also the ratio of the absolute value of their GVMs reveals the ratio of amplitude of corresponding thermal tunability.
The above simulation results are in good agreement with our theoretical prediction, validating the relationship between GVM and thermal tunability. 

By using a thicker layer of lithium niobate ($\sim$700 nm), we are able to achieve small and positive GVM ($<$200 fs/mm) to obtain larger thermal tunability. With the waveguide geometry of 700$\times$1500 nm, the required poling period is 55$\%$ larger than the thinner one, which reduces the fabrication difficulty. Additionally, better mode confinement in thicker nanowaveguide promises lower optical propagation loss, which is important for many practical applications. Thus in the following sections, we will focus on the Waveguide 2 configuration, while keeping in mind that the same technique is applicable to both.

\subsection{Fabrication}

\begin{figure}
\centering
\includegraphics[width=3.4in]{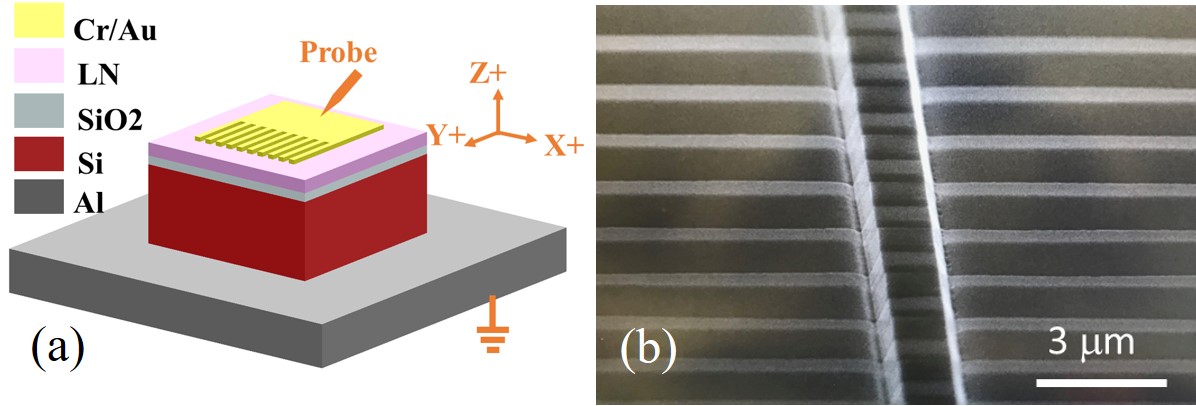}
\caption{(a) Schematic of periodic poling on the Z-cut LNOI wafer placed on an aluminum hot plate. (b) SEM image of a PPLN waveguide with 3.8 $\mu m$ poling period. The fringe pattern is formed after hydrofluoric (HF) acid etching and indicates good uniformity along the propagation direction.} \label{fig3}
\end{figure}

The PPLN waveguides are fabricated on an Z-cut thin film LNOI wafer (NANOLN Inc.), which is a 700 nm thick LN thin film bonded on a 2-$\mu m$ thermally grown silicon dioxide layer above a silicon substrate, shown in Fig.\ref{fig3} (a). We use bi-layer electron-beam resist (495 PMMA A4 and 950 PMMA A4) and define the poling electrodes by using electron-beam lithography (EBL, Elionix ELS-G100, 100 keV). Then 30-nm Cr and 60-nm Au layers are deposited via electron-beam evaporation (AJA e-beam evaporator). The desirable comb-like poling electrode pattern (see metal layer in Fig.\ref{fig3} (a)) is then created by a metal lift-off process. We apply several 1-ms high voltage ($\sim$550 V) electrical pulses on the poling pads to form the domain inversion region. The whole sample is placed on high temperature ($\sim 300 ^\circ C$) heater. Elevated temperature is a critical factor in order to reduce the required corrective voltage for thin-film lithium niobate.Then a second EBL is carried out to define the LN waveguide in the poled region. Using a similar process described in our previous work \cite{Chen:18}, an optimized ion milling process is used to shallowly etch the waveguide structure ($\sim$ 480 nm) with smooth sidewalls and the optimum sidewall angle. RCA ( (5:1:1, deionized water, ammonia and hydrogen peroxide) bath for the removal of the redeposition is applied delicately to minimize the sidewall roughness due to the uneven removal rates for poled and unpoled regions. Later on, we apply an overclad layer of 2 $\mu m$ silicon oxide on the chip with via Plasma-enhanced chemical vapor deposition (PECVD). To examine the poling quality, we use hydrofluoric (HF) acid to attack the etched waveguide and check the poling pattern under scanning electron microscope (SEM). As shown in Fig.\ref{fig3} (b), we obtain uniform periodic domain inversion along light propagation direction with such short poling period 3.8 $\mu m$. Also, we achieve good poling uniformity in thickness (down to 480 nm in depth), which is quite challenging for thick ($>$ 400 nm) lithium niobate. Because the decay rate of poling electrical field in thin-film lithium niobate is much faster than traditional bulky lithium niobate case, especially in the case that no ground layer is directly attached to lithium niobate layer.

\begin{figure}
  \centering
\includegraphics[width=3.4in]{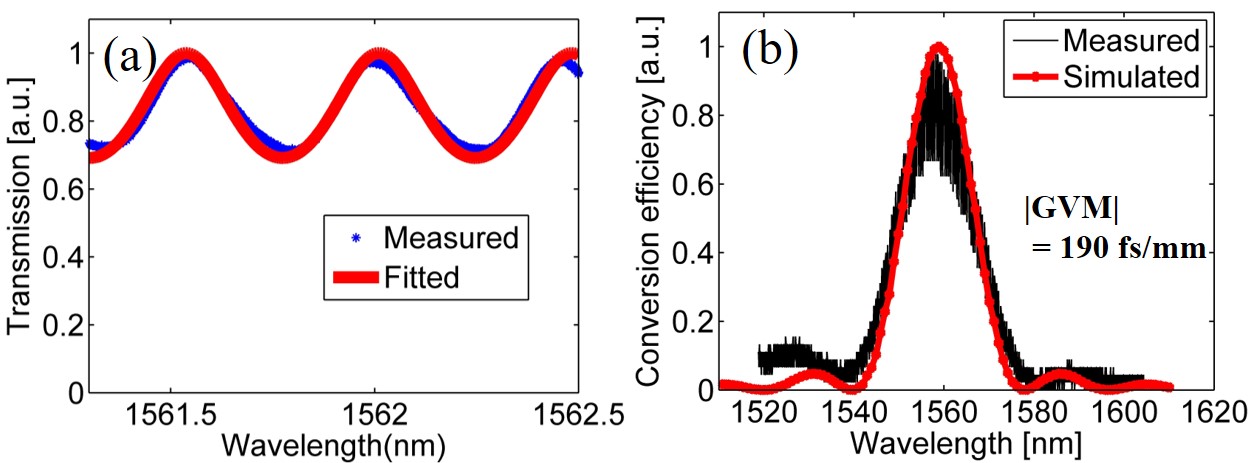}
\caption{(a) The spectra of Fabry-Perot resonances formed by the LNOI wavguide and its two polished facets. (b) Measured phase matching curve at T = $45^\circ$C. The extracted GVM magnitude is 190 fs/mm. The device length is 1-mm long.}
\label{fig4}
\end{figure}

\section{Experimental Results}
In this section, we first characterize the linear optical properties of the fabricated 1-mm long PPLN nanowaveguide. A continuous-wave (CW) tunable laser (Santec 550, 1500-1630 nm) is used as the pump to excite the waveguide's fundamental quasi-TM mode through a fiber polarization controller. Two tapered fibers (2 $\mu$m spot diameter, by OZ Optics) serve for the input/output coupling. The coupling loss at 1550 nm and 775 nm are measured to be 5.4$\pm$0.3 dB and 4.7$\pm$0.3 dB, respectively. By using the Fabry-Perot (F-P) method described in \cite{Chen:19}, the propagation loss for 1562 nm is extracted to be about 2 dB/cm. Its F-P fringes formed by the two waveguide facets are plotted in Fig.~\ref{fig4} (a). Currently, the propagation loss is primarily attributed to the surface roughness induced by the direct contact of the metal pads for poling. It can be reduced to $<0.3 $ dB/cm by avoiding the metal contact, by inserting a buffer layer in between lithium niobate layer and the top metal layer. 

Next, we characterize the waveguide's nonlinear property and measure the phase matching curve for SHG. Using the similar setup as in the above, we sweep the wavelength of the telecom laser, and collect and measure the generated second-harmonic light in the visible band with a power meter. As shown in Fig.~\ref{fig4}(b), a good $sinc^2$-like curve is measured with minimal side peaks, which indicates high-quality poling. The highest efficiency is measured up to 2400 $\%$ W$^{-1}$cm$^{-2}$, and the bandwidth is about 18 nm. From the latter, the absolute value of GVM is estimated to be 190 fs/mm, which is in agreement with the previous Lumerical simulation results.

\begin{figure}
  \centering
\includegraphics[width=3.4in]{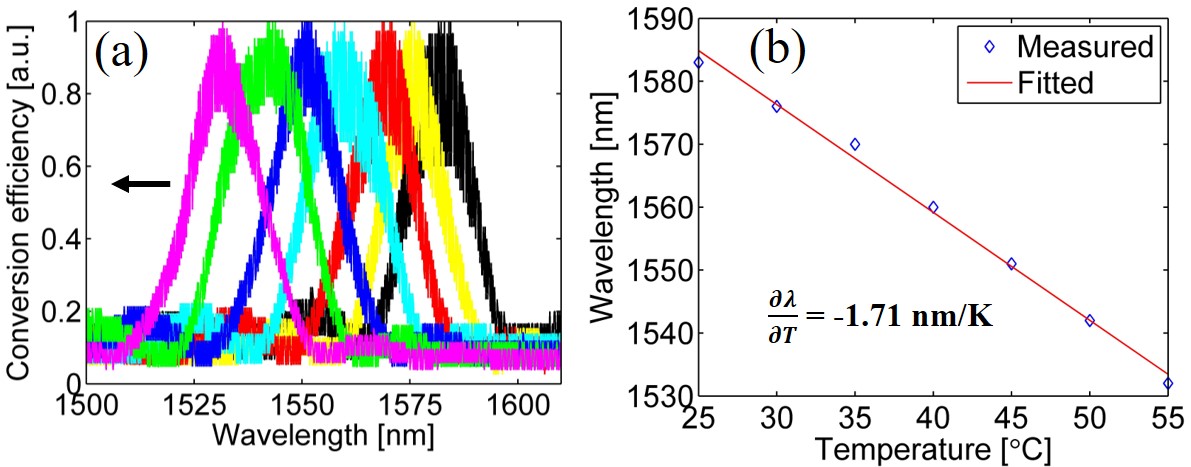}
\caption{(a) Measured phase matching curves of the LNOI waveguide blue shifted as the temperature increases from 25$^\circ$C to 55$^\circ$C at 5 $^\circ$C step. (b) Temperature dependency of the QPM wavelength, with the tunability fitted to be -1.71 nm/K.}
\label{fig5}
\end{figure}

Finally, we study the waveguide's thermal tunability. Figure.~\ref{fig5}(a) and (b) plot the phase matching profiles and their peak wavelengths measured at different temperatures. As shown, the phase matching blue shifts with increasing temperature at a linear rate of $1.71$ nm/K. To the best of our knowledge, such  blue shifting, representing a negative temperature dependency of the phase matching, is quite distinct to bulky LN waveguides or other LNOI waveguides reported so far \cite{Sua:18, luo2018highly, Wang:18PPLN, Chen:19,Elkus:19, Rao:19, niu2020optimizing}. To understand this result, we note that in Eq.\ref{eq5}, the numerator $\frac{\partial n_{775}}{\partial T} - \frac{\partial n_{1550}}{\partial T}$ is always positive. For efficient QPM, one can either use $d_{31}$ or $d_{33}$, i.e. 1550-nm o-polarized and 775-nm e-polarized for type-I or 1550-nm e-polarized and 775-nm e-polarized for type-0. For LN, that $\frac{\partial n_{e,775}}{\partial T} > \frac{\partial n_{e,1550}}{\partial T} > \frac{\partial n_{o,1550}}{\partial T}$ guarantees the positivity for the above two cases \cite{TO2005}. Then the negative temperature dependency can only come from the sign of GVM. In bulky LN or other LNOI waveguides, the GVM is always negative because of the material dispersion, thus they show positive temperature dependency in wavelength (thus negative on frequency). 

Figure~\ref{fig5}(a) also shows excellent consistency from 1530 to 1583 nm with $\pm$ 500$\%$ variation in efficiency, attributed mainly to the coupling instability. Meanwhile, the bandwidth varies within $\pm$ 1.5 nm due to material inhomogeneity or imperfect poling. We notice that the measured tunability ($\sim$ 1.71 nm/K) is almost 3 times larger than the simulated result ($\sim$ 0.6 nm/K), as shown in Figure.~\ref{fig5}(b). In addition to the thermo-optic effect, the waveguide also experiences various other effects, such as the thermal expansion and pyroelectric effect, which are not considering in the current model. Those effects are even more significant in our wavelength-scale waveguides since the optical modes are much tighter and sensitive to such environment changes. We will need to investigate in our future work.
 
Efficient, broadband, and with wide tuning range and large tuning capability, our technique is promising for frequency conversion of ultra-short coherent optical pulses, supercontinuum generation \cite{Jankowski:20}, mode selective frequency conversion \cite{Shahverdi:18} and high-dimensional quantum information processing \cite{Ansari:18}. Currently, the tuning range is limited only by the external temperature controller, which can be replaced by an on-chip microheater \cite{Lee:17} to provide an even larger tuning range.

\section{Conclusion}
In summary, we have studied group velocity engineering as a robust tool to realize $\chi^{(2)}$ processes with both high efficiency and large thermal tunability in Z-cut, periodically-poled LNOI waveguides. As a case study, we have demonstrated second harmonic generation on chip at 1900$\pm$500 $\%$W$^{-1}$cm$^{-2}$ and with the phase matching wavelength thermally tuned at -1.71 nm/K. This exceptional tunability, obtained without degrading the conversion efficiency or bandwidth, is valuable in offsetting inevitable nanofabrication errors of $\chi^{(2)}$ circuits, especially as they are mass integrated on chip. Among other applications, our technique holds the potential for the coherent light generation from visible to mid-IR spectra, providing both high power efficiency and wavelength tunability. Furthermore, with broadband (about 18 nm) phase matching in well maintained profiles over 50 nm of thermal tuning, our chips are suitable for applications ultra-short pulse. Finally, the type-0 periodic poling technique demonstrated in the present Z-cut LNOI waveguides could potentially be applied to LNOI microring cavities for significantly enhanced frequency conversion efficiency towards single photon nonlinearity \cite{chen2017observation, Chen:19optica, Lu:19}. 


{\bf Acknowledgement:} The research was supported in part by National Science Foundation (Award \#1641094 \& \#1842680) and National Aeronautics and Space Administration (Grant Number 80NSSC19K1618). Device fabrication was performed at Advanced Science Research Center, City University of New York.

{\bf Disclosures:} The authors declare that there are no conflicts of interest related to this Letter.

\bibliography{sample}

\bibliographyfullrefs{sample}


\end{document}